\documentclass[aps,prl,twocolumn,showpacs,superscriptaddress]{revtex4-1}
\usepackage{amsmath}
\usepackage{graphicx}
\usepackage{subfigure}
\usepackage{color,ulem}
\usepackage{textcomp}

\begin{document}
\title{A theory on skyrmion size and profile}
\author{X. S. Wang}
\affiliation{School of Microelectronics and Solid-State Electronics,
University of Electronic Science and Technology of China, Chengdu,
Sichuan 610054, China}
\affiliation{Physics Department, The Hong Kong University of Science and
Technology, Clear Water Bay, Kowloon, Hong Kong}
\author{H. Y. Yuan}
\affiliation{Department of Physics, Southern University of Science and
Technology, Shenzhen 518055, China}
\author{X. R. Wang}
\email{[Corresponding author:]phxwan@ust.hk}
\affiliation{Physics Department, The Hong Kong University of Science and
Technology, Clear Water Bay, Kowloon, Hong Kong}
\affiliation{HKUST Shenzhen Research Institute, Shenzhen 518057, China}
\date{\today}
\begin{abstract}
A magnetic skyrmion is a topological object consisting of an inner
domain, an outer domain, and a wall that separates the two domains.
The skyrmion size and wall width are two fundamental quantities of
a skyrmion that depend sensitively on material parameters such as
exchange energy, magnetic anisotropy, Dzyaloshinskii-Moriya
interaction, and magnetic field. However, there is no quantitative
understanding of the two quantities so far. Here, we present general
expressions for the skyrmion size and wall width obtained from energy
considerations. The two formulas agree almost perfectly with
simulations and experiments for a wide range of parameters,
including all existing materials that support skyrmions.
Furthermore, it is found that skyrmion profiles agree very well
with the Walker-like 360\textdegree{} domain wall formula.
\end{abstract}
\maketitle
	
Skyrmions, topological objects originally used to describe resonance
states of baryons \cite{skyrme}, were observed in magnetic systems
that involve Dzyaloshinskii-Moriya interaction (DMI)
\cite{Robler,Muhlbauer,Yu,Onose,Park,Heinze,Romming,Li,Jiang,news,Woo,Tian}.
There are two topologically equivalent magnetic skyrmions.
One is the Bloch skyrmions (also known as vortex skyrmions) often
found in systems with the bulk DMI \cite{Muhlbauer,Yu,Onose,Park,Tian}.
The other is the N\'{e}el skyrmions (also known as hedgehog skyrmions)
in systems with interfacial DMI \cite{Heinze,Romming,Jiang}.
Due to their small size (1 nm-100 nm) and low driven current density
(order of $10^{6}\ \mathrm{A/m^2}$) in comparison with order of $10^{12}
\ \mathrm{A/m^2}$ for a magnetic domain wall \cite{Iwasaki}, magnetic
skyrmions are believed to be potential information carriers in future
high density data storage and information processing devices
\cite{Robler,Muhlbauer,Yu,Heinze,Onose,Romming,Li,Park,Jiang,Tian,
Iwasaki,Nagaosa,Fert}.
	
Although much knowledge about magnetic skyrmions has been accumulated
after intensive studies including skyrmion generation \cite{ZhouYan,Xu,Li,
Jiang} and manipulation \cite{Yuan,Yuan2,Iwasaki,temp_grad},
the dependence of skyrmion size ($R$) on material parameters such
as exchange energy, magnetic anisotropy energy, and DMI strength is
still poorly understood at a quantitative, or even qualitative level.
A well-known formula of skyrmion size is $R \propto D/A$, where $D$
is the DMI strength and $A$ is the exchange stiffness \cite{Iwasaki}.
This formula cannot be correct because it does not capture the facts
that the skyrmion size depends sensitively on the magnetic anisotropy
$K$ \cite{Fert,MnSi_anis} and perpendicular external magnetic field
$B$ \cite{Romming,size2015}. Another formula of $R$ proposed in Ref.
\cite{skyr_trans}, which depends on $D$, $K$, and $B$, is incorrect
because skyrmion size is sensitive to the exchange stiffness $A$.
There are also other expressions for skyrmion size based on different
ansatz about the skyrmion profile \cite{JMMM1,JMMM2,Lenov2016}. However,
none of them agrees with experiments or micromagnetic simulations.
Even worse, the physical pictures behind these expressions are either
not clear or wrong. In this paper, we show that the skyrmion profiles
agree well with Walker-like 360\textdegree{} domain wall formula.
By minimizing the energy, we obtain the analytical expressions of
the skyrmion size $R$ and wall width $w$ as functions of $A$, $D$,
$K$, and $B$. Interestingly, $R$ decreases with $A$ while $w$ is
insensitive to $A$. In general, both $A$ and $w$ increases with $D$.
These results agree perfectly with micromagnetic simulations and are
consistent with experiments although they are against one's intuition.
\begin{figure}
\centering
\includegraphics[width=8.5cm]{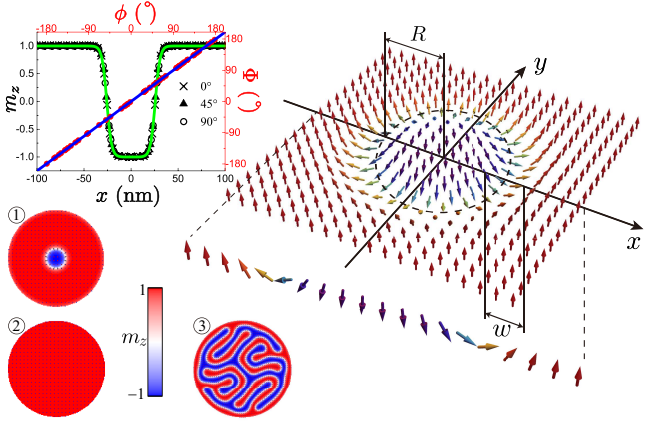}\\
\caption{Schematic diagram of a N\'{e}el skyrmion of radius $R$ and
wall (between inner and outer domains) width $w$ in a 2D film.
The arrows denote the spin direction. Spin orientations of the
skyrmion along $x$ axis is sketched below the main figure.
Upper left inset: (Right axis) $\phi$-dependence of $\Phi$
(red circles) and (Left axis) radial ($r$) distribution of
$m_z=\cos\Theta$ along the diameters of $\phi=0$ (crosses),
$45^\circ$ (triangles), and $90^\circ$ (circles) for
$A=15$ pJ/m, $D=3.7$ mJ/m$^2$, $K_u=0.8$ MJ/m$^3$, and $B=0$.
The green solid line is the fit to $\Theta_\mathrm{dw}(r)=
2\arctan\left[\frac{\sinh(R/w)}{\sinh(r/w)}\right]$ with fitting
parameters of $R=25.77$ nm and $w=4.94$ nm. The blue solid line is
$\Phi=\phi$. Lower left inset: Three typical equilibrium states
obtained in the numerical simulations for different parameters.
\textcircled{1}: isolated skyrmion. \textcircled{2}: single-domain
state of $m_z=1$ (or $ m_z=-1$). 
\textcircled{3}: stripe domains. The pixel color
encodes the $m_z$ component with the color bar shown in the figure.
}
\label{Fig1}
\end{figure}
	
We consider a two-dimensional (2D) ferromagnetic film in $xy$ plane
with an exchange constant $A$, an interfacial DMI coefficient $D$,
a perpendicular easy-axis anisotropy $K$, and a perpendicular
magnetic field $B$. The total energy $E$ of the system consists of
the exchange energy $E_\mathrm{ex}$, the DMI energy $E_\mathrm{DM}$,
the anisotropy energy $E_\mathrm{an}$, and the Zeeman energy
$E_\mathrm{Ze}$,
\begin{equation}
E=E_\mathrm{ex}+E_\mathrm{DM}+E_\mathrm{an}+E_\mathrm{Ze},
\label{energy}
\end{equation}
where $E_\mathrm{ex}=A\iint |\nabla \mathbf{m}|^2\mathrm{d}S$,
$E_\mathrm{DM}=D\iint [m_z\nabla\cdot\mathbf{m}(\mathbf{m}\cdot
\nabla)m_z]\mathrm{d}S$, $E_\mathrm{an}=K\iint (1-m_z^2)\mathrm{d}S$,
and $E_\mathrm{Ze}=\mu_0M_sB\iint (1-m_z)\mathrm{d}S$. $\mathbf{m}$
is the unit vector of magnetization of a constant saturation
magnetization $M_s$ and the integration is over the whole film.
The energy reference is chosen in such a way that the energy of
single domain state of $m_z=1$ is $E=0$. The demagnetization energy
is included in $E_\mathrm{an}$ by using the effective anisotropy
$K=K_u-\mu_0M_s^2/2$ corrected by the shape anisotropy, where
$K_u$ is the perpendicular magnetocrystalline anisotropy.
It is convenient to use a polar coordinates so that a point
$\mathbf{r}$ in the plane is denoted by $r$ and $\phi$.
Magnetization at $\mathbf{r}$ is described by polar and
azimuthal angles $\Theta(r,\phi)$ and $\Phi(r,\phi)$ so that
$\mathbf{m}=(\sin\Theta \cos\Phi,\sin\Theta \sin\Phi,\cos\Theta)$.
A skyrmion centered at $r=0$ can be described \cite{Nagaosa} by,
\begin{equation}
\Theta=\Theta(r),\quad \Phi=v\phi+\gamma,
\label{profile1}
\end{equation}
with boundary conditions of $\Theta(0)=0 (\pi)$ and $\Theta(\infty)
=\pi (0)$. $v$ is the vorticity ($v=1$ for a skyrmion and $v=-1$
for an antiskyrmion), and $\gamma$ is a constant classifying type
of skyrmions ($\gamma=0$ or $\pi$ for N\'{e}el skyrmions and
$\gamma=\pm\pi/2$ for Bloch skyrmions). A skyrmion consists of an
inner domain, an outer domain, and a wall separating the two domains.
We define the skyrmion size $R$ as the radius of the $m_z=0$ contour.
The wall width $w$ is another fundamental skyrmion quantity
often ignored in previous studies \cite{JMMM1,JMMM2,skyr_trans}.
One can also define the skyrmion polarity as $p=[m_z(r=\infty)-
m_z(r=0)]/2$ so that $p=1$ ($-1$), corresponds to spins in the
inner and outer domains pointing respectively to the $-z$($+z$)
and $+z$($-z$)-directions.

In terms of $\Theta$, four energy terms for $v=1$ are
\begin{equation}
\begin{gathered}
E_\mathrm{ex}=2\pi A\int_0^\infty \left[\left(\frac{\mathrm{d}\Theta}{\mathrm{d}r}\right)
+\frac{\sin^2\Theta}{r^2}\right]r\mathrm{d}r,\\
E_\mathrm{DM}=2\pi D \cos\gamma\int_0^\infty\left(\frac{\mathrm{d}\Theta}{\mathrm{d}r}
+\frac{\sin2\Theta}{2r}\right)r\mathrm{d}r,\\
E_\mathrm{an}=2\pi K \int_0^\infty \sin^2\Theta r\mathrm{d}r,\\
E_\mathrm{Ze}=2\pi \mu_0M_sB \int_0^\infty(1-\cos\Theta) r\mathrm{d}r.\\
\end{gathered}\label{energies}
\end{equation}
For a skyrmion of $p=1$ as shown in Fig. 1,
$\int_0^\infty\frac{\mathrm{d}\Theta}{\mathrm{d}r}r\mathrm{d}r<0$
because $\Theta$ decreases monotonically from $\Theta(0)=\pi$ to
$\Theta(\infty)=0$.
$\int_0^\infty\frac{\sin2\Theta}{2r}r\mathrm{d}r=\int_0^\infty\sin\Theta
\cos\Theta\mathrm{d}r<0$ is small because $\sin\Theta\cos\Theta=0$
far from the skyrmion wall and $\sin\Theta\cos\Theta$ changes its sign
from negative near the inner domain to positive near the outer domain.
Thus, to lower the total energy, one needs $\gamma=0 (\pi)$ for $D>0$
($<0$). This corresponds to a N\'{e}el skyrmion.
Along a radial direction, the magnetization profile looks like a
360\textdegree{} N\'{e}el domain wall as illustrated in Fig. 1.
This leads us to model a skyrmion profile by the Walker-like
360\textdegree{} domain wall solution \cite{size2015,size2016,Braun},
\begin{equation}
\Theta_\mathrm{dw}(r)=2\arctan \left[\frac{\sinh(R/w)}{\sinh(r/w)}\right].
\label{360DW}
\end{equation}

To test how good ansatz \eqref{360DW} is for a skyrmion, we use MuMax3
\cite{MuMax3} to simulate various magnetic stable states in a
magnetic disk of diameter 512 nm and thickness 0.4 nm. The mesh size
of 1 nm $\times$ 1 nm $\times$ 0.4 nm is used in our simulations.
$A=15$ pJ/m, $M_s=580$ kA/m, and perpendicular easy-axis anisotropy
$K_u=0.8$ MJ/m$^3$ \cite{Fert} are used to mimic Co layer in
Pt/Co/MgO system. The initial state is $m_z=1$ for $r>10$ nm and
$m_z=-1$ for $r\leq 10$ nm. The final stable state depends on the
values of $D$ and $B$. The lower left inset of Fig. 1 is three
typical stable states. \textcircled{1} is a skyrmion for
$D=3.7$ mJ/m$^2$ and $B=0$. \textcircled{2} is a single-domain state
of $m_z=1$ (or $m_z=-1$) for $D=0$ and $B=0$. \textcircled{3} is a
stripe domains state for $D=5$ mJ/m$^2$ and $B=0$.
The upper left plot of Fig. 1 shows the spatial distribution of $m_z$
of the skyrmion in \textcircled{1} along three radial directions,
$\phi=0$ (crosses), $45^\circ$ (triangles), and $90^\circ$ (circles).
All three sets of data are on the same smooth curve, showing $m_z$
is a function of $r$, but not $\phi$. The curve can fit perfectly
to Eq. \eqref{360DW} with $R=25.77$ nm, $w=4.94$ nm. We plotted also
$\Phi(\phi)$ at randomly picked spins from the simulated skyrmion.
All numerical data (red circles) are perfectly on $\Phi=\phi$.
These results not only confirm the validity of skyrmion expression of
Eq. \eqref{profile1}, but also suggest that $m_z(r)\equiv \cos
[\Theta_\mathrm{dw}(r)]$ follows the Walker-like 360\textdegree{}
DW profile \eqref{360DW}.
\begin{figure}
\centering
\includegraphics[width=8.5cm]{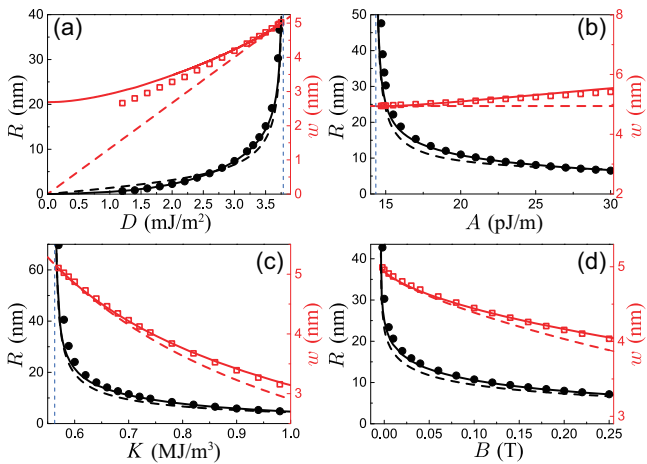}\\
\caption{The $D$ (a), $A$ (b), $K$ (c) and $B$ dependences of
skyrmion size $R$ (left axis) and wall width $w$ (right axis).
Model parameters are $A=15$ pJ/m, $K=K_u-\mu_0M_s^2=0.588$ MJ/m$^3$,
$D=3.7$ mJ/m$^2$, and $B=0$. In each subfigure, one of these four
parameters is treated as a tuning parameter, and the other three
parameters are fixed to above values. The symbols are the
micromagnetic simulation data. The solid lines are exact
analytical results obtained from Eq. \eqref{analytical}.
The dashed lines are approximate results of $R=\pi D\sqrt{\frac{A}
{16AK^2-\pi^2D^2K}}$, $w=\frac{\pi D}{4K}$ [subfigures (a)-(c)]
and solution of Eq. \eqref{min1}\eqref{min2} [subfigure (d)].
Vertical dashed lines are the upper (lower) bound of parameters
above (below) which a stable skyrmion cannot exist. }
\label{Fig2}
\end{figure}

The energy of a skyrmion can then be obtained from Eq.
\eqref{energies} by using the Walker-like 360\textdegree{}
domain wall profile $\Theta_\mathrm{dw}(r)$.
The total energy is, in general, a function of $R$ and $w$
[instead of a functional of $\Theta(r)$ and $\Phi(\phi)$] as
\begin{multline}
E(R,w)=4\pi\bigg\{A \left[f_1\left(\frac{R}{w}\right)+f_2\left(
\frac{R}{w}\right)\right]\\+Dw\left[f_3\left(\frac{R}{w}\right)
+f_4\left(\frac{R}{w}\right)\right]\\+K w^2 f_5\left(\frac{R}{w}
\right)+B w^2 f_6\left(\frac{R}{w}\right)\bigg\},
\label{analytical}
\end{multline}
where $f_i(x)$ ($i=1\sim 6$) are non-elementary functions defined
in the Supplemental Material \cite{Suppl}. The skyrmion size
and wall width $R$ and $w$ are the values that minimize $E(R,w)$.
Figure 2 are $D-$ (a), $A-$ (b), $K-$ (c) and $B-$ (d) dependences of
skyrmion size $R$ (left $y$-axis) and wall width $w$ (right $y$-axis),
with other parameters fixed to the values for Co mentioned earlier.
The symbols are the micromagnetic simulation data [$R$ is the size of
$m_z=0$ contour line and $w$ is the fit of skyrmion profile to
$\Theta_\mathrm{dw}(r)$].
Solid lines are numerical results from Eq. \eqref{analytical}.
The simulation results agree almost perfectly (except slight
deviation in the $D$-dependence of $w$ for smaller $D$) with our
analytical results of Eq. \eqref{analytical}.
Both micromagnetic simulations and analytical results clearly show
that skyrmion can exist for $D <3.8$ mJ/m$^2$ in the current case.
Above the upper limit, the stable state is not a skyrmion, but
stripe domains as shown in \textcircled{3} for $D=5$ mJ/m$^2$.
$E(R,w)$ of \eqref{analytical} has a minimum as long as $|D|<3.8$
mJ/m$^2$ that indicates existence of skyrmion.
However, micromagnetic simulation shows that skyrmion can
only exist in the window of $1.2$ mJ/m$^2< D <3.8$ mJ/m$^2$ when
the skyrmion size is larger than 1 nm in the current case.
Below 1.2 mJ/m$^2$, the stable state is a single domain with
all spins pointing up or down as shown in \textcircled{2} of Fig. 1.
This discrepancy may be due to the discretization of continuous LLG
equation in micromagnetic simulation. In principle, the mesh size
should be much smaller than the skyrmion size. For a skyrmion of 1
nm, our mesh size is 0.01 nm. Due to the limited precision of the
MuMax3 package, the mesh size cannot be further decreased.
There exists also a minimal $A$ of around 14 pJ/m as shown in Fig. 2(b)
and a minimal $K$ of around $0.56~\mathrm{MJ/m^3}$ as shown in Fig.
2(c),  below which skyrmion does not exist, and the stable state is
stripe domains as shown in \textcircled{3} of Fig. 1.
The skyrmion size decreases with $B$, which is consistent with
the experimental observations \cite{Romming,size2015,Tian}.

It is still unclear how $R$ and $w$ depend on $A$, $D$ and $K$ although
Eq. \eqref{analytical} agrees almost perfectly with simulation results.
Thus it is highly desirable to have a simple approximate expressions
for $R$ and $w$ in terms of material parameters. The exchange and DMI
energies come from the spatial magnetization variation rate.
For a skyrmion, the magnetization variation rates in the radial
and tangent directions scale respectively as $1/w$ and $1/R$.
The exchange energy is then proportional to skyrmion wall area of
$\pi Rw$ multiplying the square of the magnetization variation
rates $1/R^2+1/w^2$, i.e. $E_\mathrm {ex}\propto (R/w+w/R)$.
For a N\'{e}el skyrmion, the magnetization variation rate along the
tangent direction is perpendicular to $\mathbf{m}$ and does not
contribute to the DMI energy. The DMI energy is then proportional
to wall area ($Rw$) multiplying the magnetization variation rate
along radial direction ($1/w$), i.e. $E_\mathrm {DM}\propto R$.
The anisotropy energy is mainly from the skyrmion wall area.
Thus, $E_\mathrm{an}\propto Rw$. The Zeeman energy of the skyrmion
comes from the inner domain proportional to its area of $\pi(R-cw)^2$,
where $c$ is a coefficient depending on the magnetization profile,
and from the wall area proportional to its area of $\pi Rw$.
To obtain the proportional coefficients, one needs to find approximate
expressions for $f_i(R/w)$ ($i=1,\ldots,6$) in Eq. \eqref{analytical}.
In the case of $R\gg w$ (or $x\equiv R/w\gg 1$), $\sinh(x)\approx\cosh(x)
\approx e^x$. Thus, function $g(t,x)=[2\sinh^2 (x)\cosh^2(t)]/
[\sinh^2(x)+\sinh^2(t)]^2\approx 2e^{2(x-t)}/[e^{2(x-t)}+1]^2$ is
positive and significantly non-zero only near $t=x$, reflecting the fact
that $E_\mathrm{ex}$, $E_\mathrm{DM}$, and $E_\mathrm{an}$ are mainly
from skyrmion wall region that is assumed to be very thin.
Furthermore, the area bounded by $g(t,x)$-curve and $t$-axis is 1 so that
$g(t,x)\approx\delta(t-x)$ resembles the properties of a delta function.

We can evaluate $f_i$'s under this approximation (See Supplemental
Materials \cite{Suppl}). For example, $f_1(x)$ is
\begin{equation}
f_1(x)=\int_0^\infty g(t,x) t\mathrm{d}t\approx \int_0^\infty
\delta(x-t)t\mathrm{d}t= x.
\end{equation}
The total energy is then
\begin{multline}\label{appr}
E(R,w)=4\pi\bigg[A\left(\frac{R}{w}+\frac{w}{R}\right)-\frac{\pi}{2}DR\\
  +KwR+\mu_0M_sB\left(\frac{R^2}{2}+\frac{\pi^2}{24}w^2\right)\bigg].
\end{multline}
Due to the specific form of the magnetization profile of $\Theta_
\mathrm{dw}(r)$, $Rw$-term in $E_\mathrm{Ze}$ vanishes and $E_\mathrm{Ze}
\approx 4\pi\mu_0M_sB\left(\frac{R^2}{2}+\frac{\pi^2}{24}w^2\right)$.
The skyrmion size and wall width are then the values that make
$E(R,w)$ minimal, or
\begin{gather}
A\left(\frac{1}{w}-\frac{w}{R^2}\right)-\frac{\pi}{2}D+Kw+\mu_0M_sBR=0, \label{min1} \\
A\left(-\frac{R}{w^2}+\frac{1}{R}\right)+KR+\frac{\pi^2}{12}\mu_0M_sBw=0. \label{min2}
\end{gather}
For $B=0$, Eqs. \eqref{min1} and \eqref{min2} can be analytically solved.
The results are
\begin{equation}
R=\pi D\sqrt{\frac{A}{16AK^2-\pi^2D^2K}},\quad w
=\frac{\pi D}{4K}.
\label{brief}
\end{equation}
The dashed lines in Fig. 2(a)-(c) are the approximate
formulas that compare quite well with simulation results too.
For $B\neq0$, it is difficult to analytically solve Eqs.
\eqref{min1} and \eqref{min2}, but their numerical solutions are
easily obtained that are plotted as dashed lines in Fig. 2(d).
In summary, our approximate formula agrees very well with the
simulations for $R\gg w$ as expected from our approximation.
For smaller skyrmions, the approximation is still not bad,
and qualitatively gives correct parameter dependence.
We can also determine the upper limit of $D$ and lower
limits of $A$, $K$, and $B$ from the approximate formula.
Since $R$ must be real and finite, we have
\begin{equation}
D<\frac{4}{\pi}\sqrt{AK},\; A>\frac{\pi^2D^2}{16K},\;
K>\frac{\pi^2D^2}{16A},
\end{equation}
for $B=0$. Note that these limits are consistent with
the criteria of the existence of chiral domain walls \cite{dzy1965,Thiaville2013}.
These critical values are plotted in Fig. 2(a)-(c) as vertical
dashed lines that agrees also well with simulations.

We compare our theoretical results of skyrmion size with
the experimental results for PdFe/Ir \cite{Romming,size2015,PdFeIr}
and W/Co$_{20}$Fe$_{60}$B$_{20}$/MgO \cite{Jiang,BandK},
in which isolated skyrmions are observed. For PdFe/Ir, the
parameters are $M_s=961\sim1160$ kA/m, $A=2\sim4.87$ pJ/m,
$K=2.5$ MJ/m$^3$, $D=3.4\sim 3.9$ mJ/m$^2$, and $B=1.15\sim2.97$
T \cite{Romming,size2015}. Our theory gives small skyrmion size
of $0.53\sim1.59$ nm that compares well with the experimental
results of $0.9\sim1.9$ nm in Ref. \cite{Romming}.
For $\mathrm{W/Co_{20}Fe_{60}B_{20}/MgO}$, the parameters are
$M_s=650$ kA/m, $A=10$ pJ/m, $K=0.02275$ MJ/m$^3$,
$D=0.68\sim 0.73$ mJ/m$^2$, and $B=0.00025\sim0.0005$ T \cite{Jiang,BandK}.
Our theory gives large skyrmion size of $356\sim1484$ nm, consistent
with the experimental results $700\sim2000$ nm in Ref. \cite{Jiang}.
Our theoretical results show good agreement with the
experiments although some of the material parameters can only be
roughly estimated from different literatures.
	
We also compare our analytical results with micromagnetic simulations for
PdFe/Ir \cite{Romming,size2015,PdFeIr},
MnSi \cite{MnSi_anis,MnSi}, and W/Co$_{20}$Fe$_{60}$B$_{20}$/MgO \cite{Jiang}.
The skyrmion sizes range from several nanometers to about 2 micrometers.
All the comparisons give quite good agreement (See Supplemental Materials
\cite{Suppl}). Our results show that skyrmion size increases with $D$,
and decreases with $A$ and $K$.
Our results also show that not only DMI, but also magnetic anisotropy
or perpendicular magnetic field is necessary for the formation of
isolated skyrmions, which is consistent with all previous experiments
and simulations \cite{Robler,Muhlbauer,Yu,Onose,Park,Heinze,
Romming,Li,Jiang,Tian,Iwasaki,Nagaosa,Fert,MnSi_anis,size2015,size2016}.

It is natural to extend our approach to Bloch skyrmions
in the systems with bulk inversion symmetry broken.
The bulk DMI energy $E_\mathrm{DM}=D\iint \mathbf{m}\cdot(\nabla
\times\mathbf{m}) dS$ can be rewritten as
\begin{equation}
E_\mathrm{DM}=2\pi D \sin\gamma\int_0^\infty\left(\frac{\mathrm{d}
\Theta}{\mathrm{d}r}+\frac{\sin2\Theta}{2r}\right)r\mathrm{d}r,
\end{equation}
where $\gamma=\pi/2$ gives minimal energy. Since all other
discussions are the same as those for N\'{e}el skyrmions, the
results about $R$ and $w$ are applicable for the Bloch skyrmions.

In conclusion, we found a single skyrmion can be well described by a
360\textdegree{} domain wall profile parametrized by two fundamental
quantities, skyrmions size and wall width.
Through the minimization of
total energy with respect to skyrmion size and wall width, analytical
formulas for skyrmion size and wall width as a function of exchange
stiffness, anisotropy coefficient, DMI strength and external field are obtained.
The formulas agree very well with simulations and experiments.

\begin{acknowledgments}
This work was supported by the National Natural Science
Foundation of China (Grant No. 11774296 and No. 61704071) as
well as Hong Kong RGC Grants No. 16300117 and No. 16301816.
X.S.W acknowledge support from UESTC and China Postdoctoral
Science Foundation (Grant No. 2017M612932).
\end{acknowledgments}

\clearpage
\onecolumngrid
\section{Supplemental Material}
\setcounter{table}{0}
\setcounter{figure}{0}
\subsection{Derivation of energy expressions}
To derive the functions $f_i(x)$ in the energy expression,
we substitute Eq. \eqref{360DW} into Eq. \eqref{energies}.
For the exchange energy, by defining $x=R/w$, $t=r/w$, we have
\begin{align*}
E_\mathrm{ex}&=2\pi A\int_0^\infty \left[\left(\frac{\mathrm{d}\Theta}{\mathrm{d}r}\right)
+\frac{\sin^2\Theta}{r^2}\right]r\mathrm{d}r\\
&=4\pi A \int_0^\infty \left\{\frac{2\sinh^2 (x) \cosh^2(t)}{\left[\sinh^2(x)+\sinh^2(t)\right]^2}t
+\frac{2\sinh^2 (x) \sinh^2(t)}{t\left[\sinh^2(x)+\sinh^2(t)\right]^2}\right\}dt
\end{align*}
Thus, we define
\begin{gather*}
f_1(x)=\int_0^\infty \frac{2\sinh^2 (x) \cosh^2(t)}{\left[\sinh^2(x)+\sinh^2(t)\right]^2}t\mathrm{d}t,\\
f_2(x)=\int_0^\infty \frac{2\sinh^2 (x) \sinh^2(t)}{t\left[\sinh^2(x)+\sinh^2(t)\right]^2}\mathrm{d}t.
\end{gather*}
While $R\gg w$ ($x\gg 1$), we have $\sinh(x)\approx\cosh(x)\approx e^x$, so that
\begin{equation*}
f_1(x)\approx\int_0^\infty \frac{2e^{2(x-t)}}{\left[e^{2(x-t)}+1\right]^2}t\mathrm{d}t\\
\end{equation*}
The function $\frac{2e^{2(x-t)}}{\left[e^{2(x-t)}+1\right]^2}$ is non-zero only for $x\approx t$.
Approximately, we have
\begin{equation}
\frac{2e^{2(x-t)}}{\left[e^{2(x-t)}+1\right]^2}\approx I_1 \delta(x-t),
\end{equation}
where the coefficient $I_1$ is determined by $I_1=\int_{-\infty}^{
\infty}\frac{2e^{2x}}{\left(e^{2x}+1\right)^2}\mathrm{d}x=1$. Thus,
$f_1(x)\approx \int_0^\infty t\delta(x-t)\mathrm{d}t=x$.
Similarly, $f_2(x)\approx \int_0^\infty \delta(x-t)/t\mathrm{d}t=1/x$.

For the DM energy, we have
\begin{align*}
E_\mathrm{DM}&=2\pi D \cos\gamma\int_0^\infty\left(\frac{\mathrm{d}\Theta}{\mathrm{d}r}
+\frac{\sin2\Theta}{2r}\right)r\mathrm{d}r\\
&=4\pi D w \int_0^\infty\left[-\frac{\sinh x\cosh t}{\sinh^2x+\sinh^2t}t-\frac{\sinh x\sinh t(\sinh^2x-
\sinh^2t)}{(\sinh^2x+\sinh^2t)^2} \right]\mathrm{d}t.
\end{align*}
We define
\begin{gather*}
f_3(x)=-\int_0^\infty \frac{t\sinh(x) \cosh(t)}{\sinh^2(x)+\sinh^2(t)}\mathrm{d}t,\\
f_4(x)=-\int_0^\infty \frac{t\sinh(x) \sinh(t)\left[\sinh^2(x)-\sinh^2(t)\right]}{\left[\sinh^2(x)+\sinh^2(t)\right]^2}\mathrm{d}t.\\
\end{gather*}
For $f_3(x)$, the function inside the integral $\frac{\sinh(x) \cosh(t)}{\sinh^2(x)+\sinh^2(t)}
\approx \frac{e^{(x-t)}}{e^{2(x-t)}+1}$ is localize at $x=t$ so that we have
the approximation
\begin{align*}
f_3(x)&\approx -\int_0^\infty \frac{e^{(x-t)}}{e^{2(x-t)}+1}t\mathrm{d}t\\
&\approx -\int_0^\infty I_3\delta(x-t)t\mathrm{d}t\\
&\approx I_3x,
\end{align*}
where $I_3$ is determined by $I_3= \int_{-\infty}^\infty \frac{e^{x}}{e^{2x}+1} \mathrm{d}x
=\pi/2$.
The integrand in $f_4$ is 0 at $r=0$, $r=\infty$ and $r=R$. Furthermore, it has opposite signs for
$r<R$ and $r>R$. So $f_4\sim 0$ after the integration.

For the anisotropy energy, we have
\begin{equation*}
f_5(x)=\int_0^\infty \frac{2t\sinh^2 (x) \sinh^2(t)}{\left[\sinh^2(x)+\sinh^2(t)\right]^2}\mathrm{d}t.
\end{equation*}
Similar to the exchange energy and DM energy, the anisotropy energy is
only non-zero near $r=R$, too. The approximate form of function $f_5$
is the same as $f_1$,
\begin{align*}
f_5(x)&\approx \int_0^\infty \frac{2t e^{2 (x-t)}}{\left[e^{2(x-t)}+1\right]^2}\mathrm{d}t\\
&\approx \int_0^\infty t \delta(x-t) \mathrm{d}t=x.
\end{align*}

The Zeeman energy is non-zero for both the wall region and the inner domain.
The function $f_6$ is
\begin{equation*}
f_6(x)=\frac{\sinh^2(x)}{\sinh^2t+\sinh^2x}t\mathrm{d}t
\end{equation*}
Again, we replace $\sinh$ functions by exponential functions to obtain
\begin{align*}
f_6(x)&\approx\int_0^\infty \frac{1}{e^{2(t-x)}+1}t\mathrm{d}t\\
&=-\frac{1}{4}\textrm{Li}_2(-e^{2x}),
\end{align*}
where $\mathrm{Li}_s(x)$ is the polylogorithm function defined by
$\mathrm{Li}_s(x)=\sum_{n=1}^\infty \frac{x^n}{n^s}$. The asymptotic
form of $-\frac{1}{4}\textrm{Li}_2(-e^{2x})$ is
\begin{equation*}
\lim_{x\rightarrow\infty}-\frac{1}{4}\textrm{Li}_2(-e^{2x})=\frac{x^2}{2}+\frac{\pi^2}{24}
\end{equation*}
So we have $f_6(x)\approx \frac{x^2}{2}+\frac{\pi^2}{24}$.

\subsection{Numerical verification of theoretical results for different materials}

We compare the theoretical results with micromagnetic simulations.
Similar to Fig. 2 in the main text, we compare the $D$-, $A$-, $K$-,
and $B$-dependencies of skyrmion size $r$ and skyrmion wall width $w$.
The sample size ranges from 256 nm to 2048 nm, and the sample
thickness is fixed to 0.1 nm.
In each subfigure, one of $D$, $A$, $K$, and $B$ is treated as a tuning
parameter, and the other three parameters are fixed to the values listed
in Table S1.
\begin{table}[!hbp]

	\renewcommand\thetable{S\arabic{table}}
	\begin{tabular}{l|c|c|c|c|c}

		\hline
		
		&$A$ (pJ/m) & $K$ (MJ/m$^3$) & $D$ (mJ/m$^2$) & $M_s$ (kA/m) &$B$ (T)\\
		\hline
		\hline
		
		PdFe/Ir (IDMI) & 4.87 & 2.5 & 3.43 & 961 & 1.15\\
		
		\hline
		MnSi (BDMI)& 0.845 & -0.0334 & 0.338 & 163 & 1\\
		\hline
		W/Co$_{20}$Fe$_{60}$B$_{20}$/MgO (IDMI) & 10 & 0.0228 & 0.7 & 650 &  0.0005\\
		\hline

	\end{tabular}
	
	\caption{Parameters used for verification of our results. IDMI (BDMI) means
the DMI is interfacial (bulk) type.}
\end{table}

\begin{figure}
\renewcommand\thefigure{S\arabic{figure}}
\centering
\includegraphics[width=\textwidth]{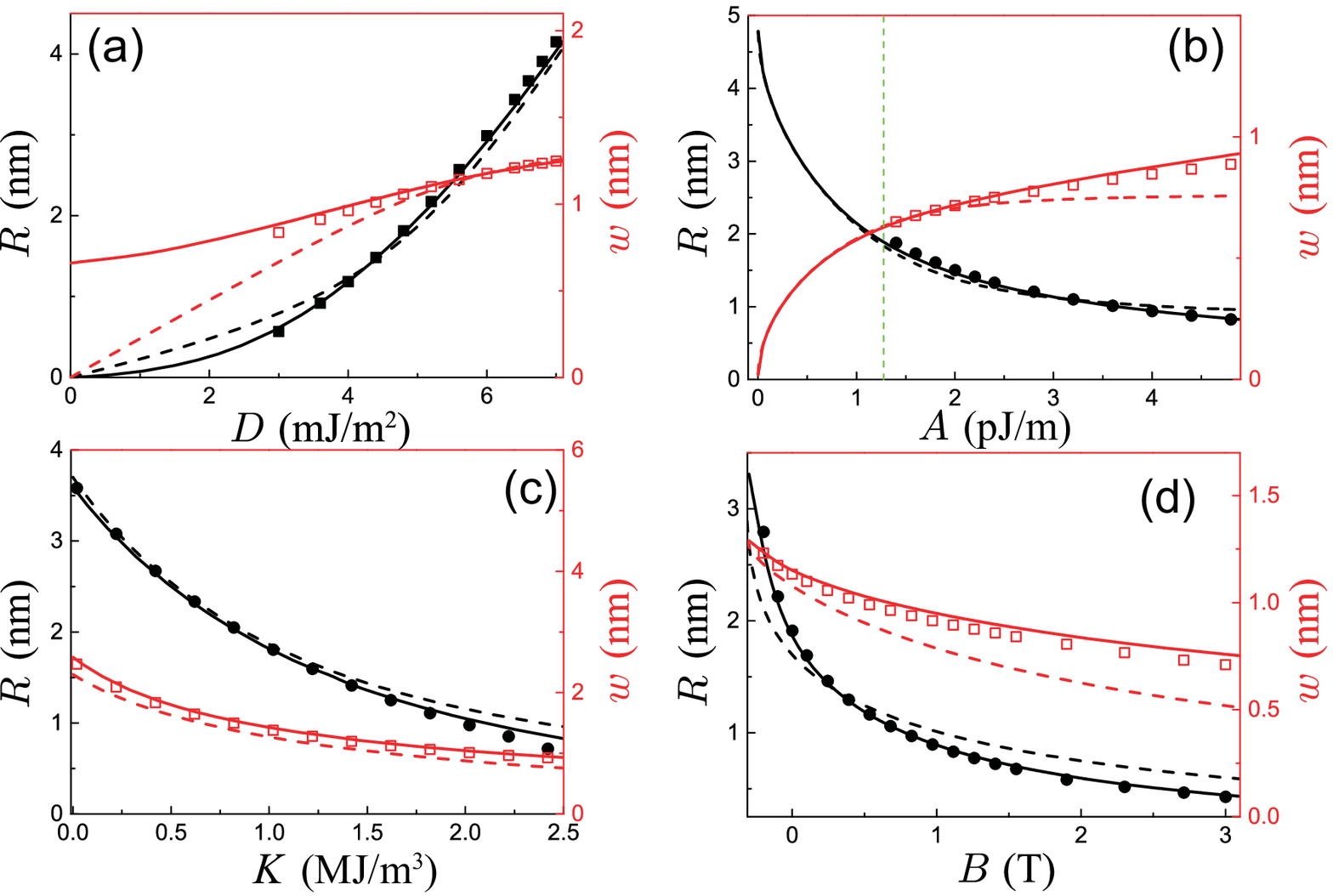}\\
\caption{Comparison between numerical and theoretical results for PdFe/Ir
parameters. The meanings of lines and symbols are the same as those in
Fig. 2 of the main text. }
\label{FigS1}
\end{figure}

\begin{figure}
\renewcommand\thefigure{S\arabic{figure}}
\centering
\includegraphics[width=\textwidth]{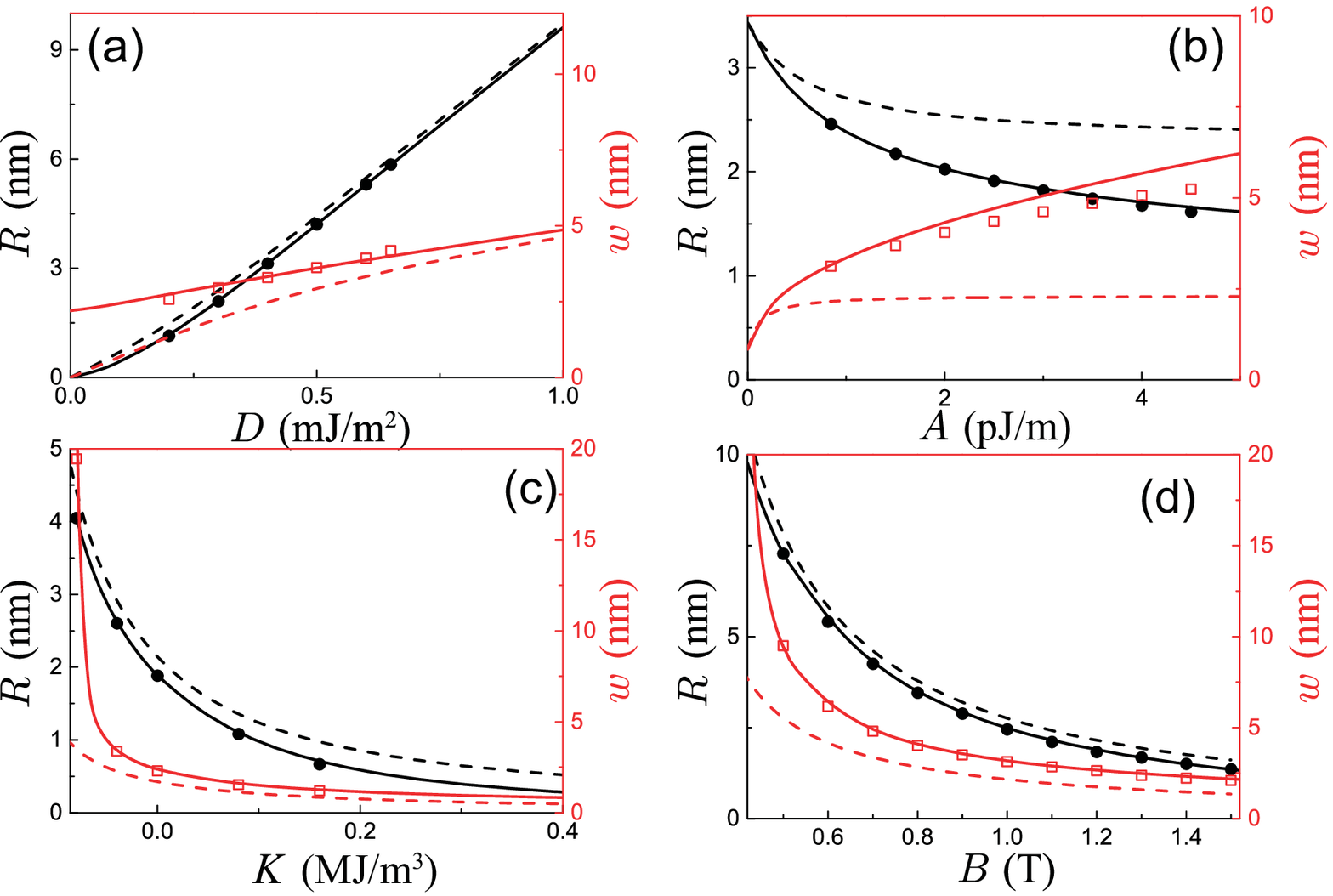}\\
\caption{Comparison between numerical and theoretical results for MnSi
parameters. The skyrmions are Bloch-type.
$R\gg w$ cannot be satisfied for parameters near the MnSi parameter,
so the approximate formulas (dashed lines) do not agree well with the exact formulas
(solid lines).
Nevertheless, the numerical simulation results agree well with the exact formulas.}
\label{FigS2}
\end{figure}
\begin{figure}
\renewcommand\thefigure{S\arabic{figure}}
\centering
\includegraphics[width=\textwidth]{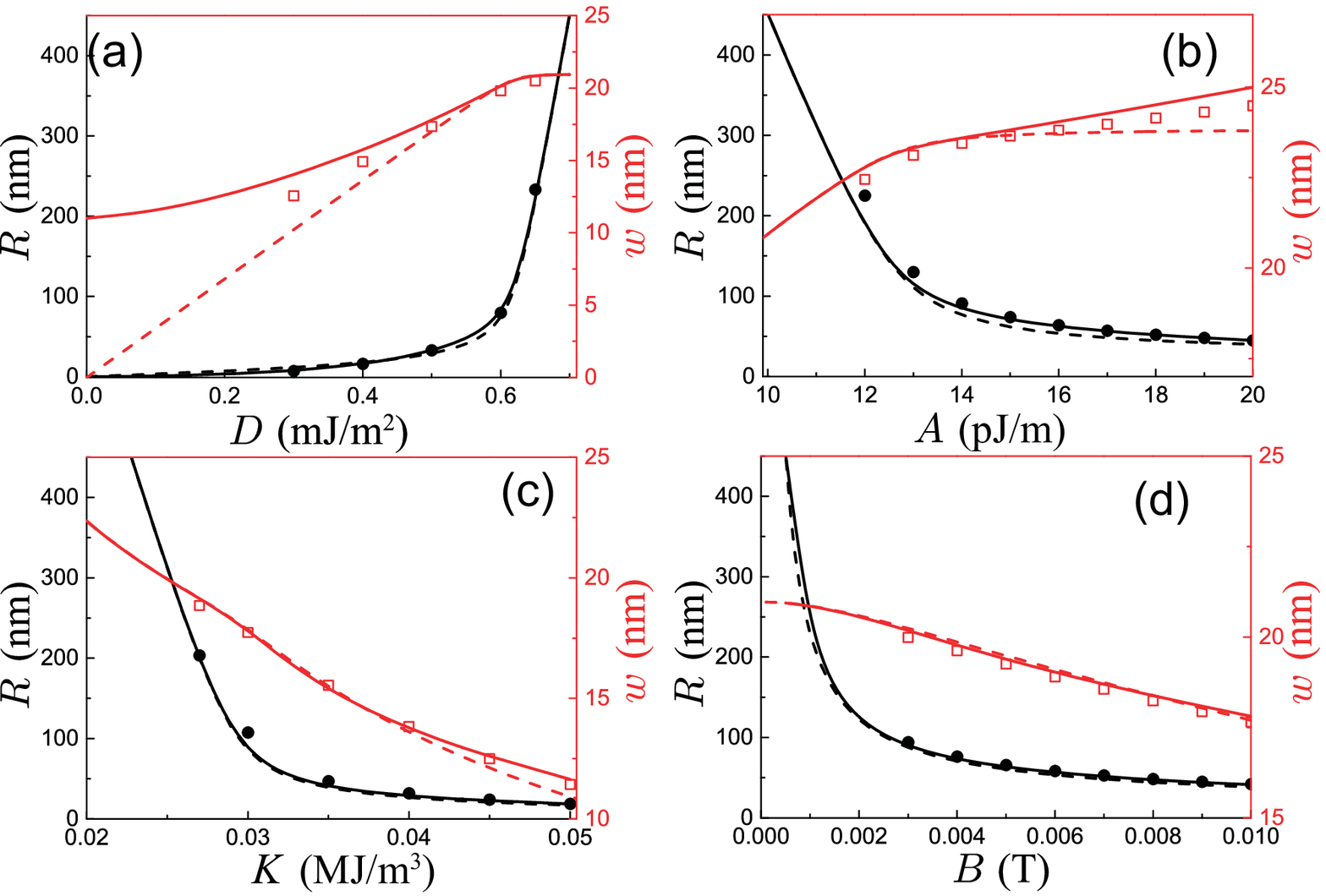}\\
\caption{Comparison between numerical and theoretical results for W/Co$_{20}$Fe$_{60}$B$_{20}$/MgO
parameters. }
\label{FigS3}
\end{figure}

\end{document}